\documentclass[final,5p,authoryear]{elsarticle}

\preprintfalse
\ifpreprint
\usepackage[nolists,noheads]{endfloat}

\fi

\usepackage[british]{babel}
\usepackage{graphicx}
\usepackage{longtable}
\usepackage{amssymb}
\usepackage{color}
\usepackage{rotating}
\usepackage{overpic}
\usepackage{hhline}
\usepackage{booktabs}

\usepackage{lineno}
\ifpreprint
\fi

\newcommand{\reftab}[1]{Table~\ref{#1}}

\newcommand{\reffig}[1]{Figure~\ref{#1}}
\newcommand{\isotope}{\ensuremath{\delta^{18}\mathrm{O}}}

\newcommand{\comment}[1]{}

\journal{Journal of Archaeological Sciences}

\begin{document}
\begin{frontmatter}

\title{On the sensitivity of the simulated European Neolithic transition to climate extremes}

\author{Carsten Lemmen\corref{lab:cor}}
\ead{carsten.lemmen@hzg.de}

\author{Kai W. Wirtz} 
\cortext[lab:cor]{Tel +49\,4152\,87-2013, Fax~-2020} 
\address{Helmholtz-Zentrum Geesthacht, Institute of Coastal Research, Max-Planck Stra\ss e~1, 21501~Geesthacht, Germany}

\begin{abstract}
Was the spread of agropastoralism from the Fertile Crescent throughout Europe influenced by extreme climate events, or was it independent of climate? We here generate idealized climate events using  palaeoclimate records. In a mathematical model of regional sociocultural development, these events disturb the subsistence base of simulated forager and farmer societies.  We evaluate the regional simulated transition timings and durations against a published large set of radiocarbon dates for western Eurasia; the model is able to realistically hindcast much of the inhomogeneous space-time evolution of regional Neolithic transitions. Our study shows that the consideration of climate events improves the simulation of typical lags between cultural complexes, but that the overall difference to a model without climate events is not significant.  Climate events may not have been as important for early sociocultural dynamics as endogenous factors. 
\end{abstract}

\end{frontmatter}

%

\section{Introduction}
%
Between 10\,000 and 3000\,cal\,BC, western Eurasia saw enormous cultural, technological, and sociopolitical changes with the emergence of agropastoralism, permanent settlements, and state formation  \citep{Barker2006}.  Human population experienced a dramatic increase \citep{BocquetAppel2008end,Gignoux2011},  and  people, plants and animals moved or were moved great distances \citep[e.g.,][]{Zohary1993}.  

While the Holocene possibly defines the start of  major anthropogenic global environmental change \citep{Lemmen2010,Kaplan2011}, it also marks the period where climatic shifts could have affected human subsistence more severely than ever before: reduced mobility after investments in settlement infrastructure most likely increased the sensitivity of the novel farmers to environmental alterations  \citep{Janssen2004}. There remains, however, considerable uncertainty on whether and how climate instabilities had influenced the development and spread of agropastoralism in Eurasia \citep{Berglund2003,Coombes2005}.

\subsection{Origin and spread of western Eurasian farming}
%
The Neolithic originated most probably in the Fertile Crescent, between the Levantine coast and the Zagros ridge.  In this region, almost all European food crops and animals---wheat, barley, cattle, sheep, pigs---had been domesticated and inserted into a broad spectrum of foraging practices during the tenth millennium cal\,BC \citep{Flannery1973,Zeder2008}.  Neolithic (farming based) life style emerged not before the 9th~millennium~BC in this core region \citep{Rosen2012}, and expanded to Cyprus by 8500\,cal\,BC \citep{Peltenburg2000}; around 7000\,cal\,BC, agropastoralism appeared on the Balkan and in Greece \citep{Perles2001}. Propagating in a generally northwestern direction, agropastoralism finally arrived after 4000 cal\,BC on the British isles and throughout northern Europe \citep{Sheridan2007}; in a western direction, the expansion proceeded fast along the Mediterranean coast to reach the Iberian peninsula  at 5600\,cal\,BC \citep{Zapata2004}.  

\subsection{Transitions and climate}
It has been argued that a precondition of agriculture was the relatively stable environment of the Holocene  \citep{Feynman2007}, and that only in this stable environment active cultivation and establishment of infrastructure such as fields and villages was favored \citep{VanderLeeuw2008}. Within its relative stability, however, the Holocene climate exhibited variability on many spatial and temporal scales with pronounced multi-centennial and millennial cycles \citep{Mayewski2004,Wanner2008}.  In addition, non-cyclic anomalies have been identified \citep[e.g.,][]{Wirtz2010}, most prominently the so-called 8.2 and 4.2~events (around 6200 and 2200\,cal\,BC, respectively, \citealt{VonGrafenstein1998,Cullen2000}). 
Although the regional scale and intensity of the 4.2~event has been strongly questioned \citep[e.g.,][]{Finne2011},
the event had evoked the formulation of hypotheses on the connection between climatic disruptions and societal collapse \citep{Weiss1993,DeMenocal2001}. Similarly, the globally documented 8.2~event has been linked to the abandonment of many settlements in the Near East and simultaneous appearance of new village structures in southeast Europe \citep{Weninger2005}.  

It might be coincidental that the 8.2 and 4.2~events define the time window of the Neolithic expansion in Europe, but the general view that environmental pressure on early Neolithic populations may have stimulated outmigration has been put forward since long \citep{Childe1942}.  \citet{Dolukhanov1973}, \citet{Gronenborn2009tcc,Gronenborn2010}, or \citet{Weninger2009} suggest that climate-induced crises may have forced early farming communities to fission and move in order to escape conflicts. \citet{Berger2009}, to the contrary, see the role of climate events rather in creating opportunities: the rapid farming expansion into the Balkan could have been stimulated by an increase of natural fires after the 8.2~event, which opened up the formerly forested landscape.  

\subsection{How sensitive was the Neolithization to climate?}
%
The relevance of climate variability and external triggers for prehistoric agricultural dynamics has been severely questioned \citep[e.g.,][]{Erickson1999,Coombes2005}. Alternative theories of the Neolithic transition underline the agency of early societies \citep{Shanks1987,Whittle2007}.  On the other hand, the development of technological, social, and cultural complexes can hardly be thought to evolve independently of their variable environments; and the spatio-temporal imprint of the Neolithization in Eurasia requires a geographic approach which resolves how people and/or goods and practices migrated over long distances. \citet{Berglund2003}, e.g., suggested a stepwise interaction between agriculture and climate but found no strong links  for northwest Europe.  

The dispersal of agriculture into Europe has long been mathematically formulated based on \citeauthor{Childe1925}'s (\citeyear{Childe1925}) observation on the spatiotemporal distribution gradient of ceramics that \citet{Ammerman1971} formulated as the  `wave of advance' model.  This simple---and also the later more advanced ones \citep{Ackland2007,Galeta2011,Davison2006}---diffusion models received support from linguistic  \citep[e.g.][]{Renfrew1987} and archaeogenetic work \citep[e.g.][]{Balaresque2010}.  The dispersal of agriculture in these models occurs concentrically, and can be modulated by topography and geography.  This dispersal model is not able to describe the inhomogeneous spatiotemporal distribution of radiocarbon dates, which are, e.g., apparent in regionally different stagnation periods \citep[`hypoth\`ese arythmique',][]{Guilaine2003,Rasse2008,Schier2009}.  

Stagnations are visible in the simulation by \citet{Lemmen2011}, who integrate endogenous regional sociocultural dynamics with the dispersal of agriculture.  Their approach connects social dynamics---as optimally evolving agents---to regionally and temporally changing environments; in addition, they account for the spatio-temporal spread of populations and 
technological traits. Their Global Land Use and technological Evolution Simulator (GLUES) has proven to produce realistic hindcasts of the origin and distribution of agropastoralism and concomitant cultures around the globe
\citep{Wirtz2003gdm,deVries2002ufu}, for Eastern North America \citep{Lemmen2012ena}, the Indus valley \citep{Lemmen2012har}, and western Eurasia \citep{Lemmen2011}. Using GLUES and a globally synchronous
climate forcing signal, \cite{Wirtz2003gdm}  found a general delay of the simulated regional Neolithic due to climate fluctuations; at a global scale, differences in hindcasted socio-cultural trajectories proved to be largely independent of temporal disruptions.  

We here use temporal disruptions that are defined as excursions of a climate variable far from the local mean climate, i.e.\ extreme climate events; we do not consider rapid climate shifts that abruptly alter the climate mean state \citep[e.g.,][]{Dakos2008}. The hypothesis that extreme climate events had significant impacts on the Neolithization of Europe is critically examined: we employ GLUES as a deductive tool to reconstruct the Neolithic transition in Europe and evaluate the simulated reconstruction against the radiocarbon record of Neolithic sites in two experiments: (1)~one including climate events, represented by a pseudo-realistic spatially resolved climate event history for the period 9500--3000\,cal\,BC; and (2)~another without climate events. 

\section{Material and Methods}

\subsection{Reconstructing climate event history}

We used a data collection of $134$ globally distributed, high-resolution ($<200$\,a) and long-term ($>4000$\,a) palaeoclimate time series collected from public archives and published literature.  The collection only contains studies where the respective authors indicated a direct relation to climate variables such a precipitation, temperature, or wind regime \citep[e.g.][]{Bond1997,Wick2003,Chapman2000,Gasse2000}.  A large part of this data set ($122$~time series) was previously analyzed by \citet{Wirtz2010} for extreme events; a complete overview of time series in this collection is provided in the supporting online material (Table~S1).  Due to the different types of proxies originating from both marine and terrestrial sites (mostly \isotope, see \reftab{tab:proxies}) the relation to climate variables is often ambiguous, also in sign. This ambiguity does not affect our analysis, as we are only interested in the spatio-temporal characterization of extreme events: a drastic excursion from a climate mean state  stressed regional habitats and human populations regardless of its direction. 

Our data set comprises 134 palaeoclimate time series, all of them long-term and high-resolution, and provides the best spatial and temporal coverage of any study we are aware of.  Previous collections used 18, 50, 60 or 80 records (\citealt{Wanner2008,Mayewski2004,Holmgren2003,Finne2011}, respectively), mostly limited to the last 6000 years.   The coverage we use here is sufficient to represent climate variability in almost all land areas of the world (with sparsest regional coverage in central Australia, Saharan Africa, and Northern-Central Eurasia),  considering the spatial coherence of climate signals within 1500\,km distance found by \citet{Wirtz2010} for their similar data set.   

\begin{table*}
\caption{Table of 26~palaeoclimate time series and associated timing of extreme climate events.
This is a subset (relevant for the Western Eurasian focus area) of the entire global collection used to generated extreme climate events; the full dataset is shown in the supplementary material Table~S1.  Abbreviations: Lk=Lake, Cv=Cave, SST=sea surface temperature, T=temperature, T7=July temperature, P=precipitation, GSD=grayscale density, Ti=Titanium content, and HSG=hematite stained glass.}
\label{tab:proxies}
\vspace*{1ex}
\ifpreprint
\hskip-0.2\hsize\includegraphics[width=1.4\hsize]{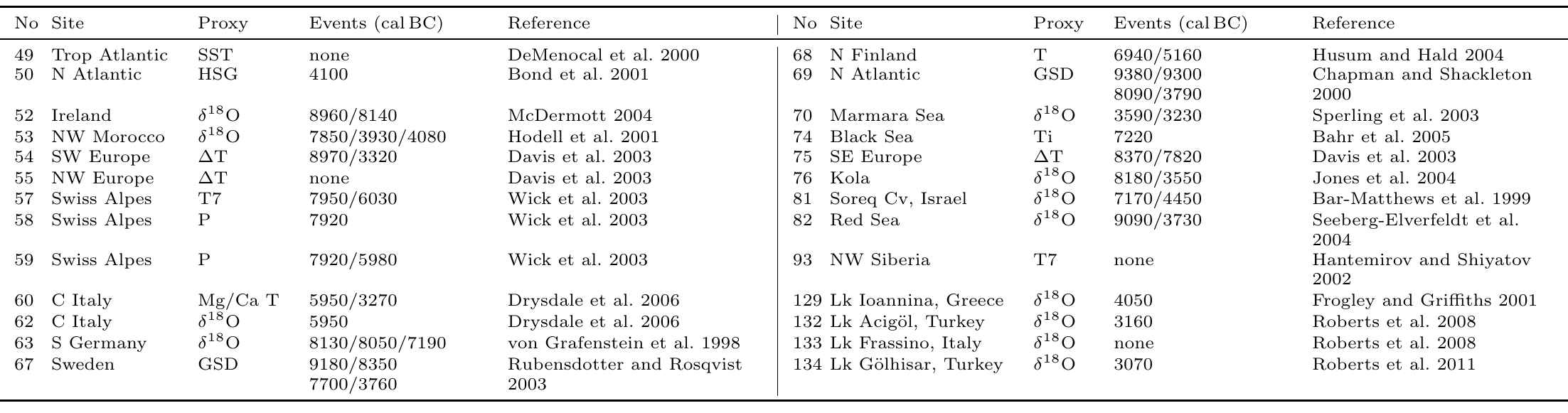}
\else
\includegraphics[width=\hsize]{table_1}
\fi
\end{table*}

\nocite{DeMenocal2000}\nocite{Husum2004}
\nocite{Bond2001}\nocite{Chapman2000}
\nocite{McDermott2004}\nocite{Sperling2003}
\nocite{Hodell2001clm}\nocite{Bahr2005}
\nocite{Davis2003}\nocite{Davis2003}
\nocite{Davis2003}\nocite{Jones2004}
\nocite{Wick2003}\nocite{BarMatthews1999}
\nocite{Wick2003}\nocite{SeebergElverfeldt2004}
\nocite{Wick2003}\nocite{Hantemirov2002}
\nocite{Drysdale2006}\nocite{Frogley2001}
\nocite{Drysdale2006}\nocite{Roberts2008}
\nocite{VonGrafenstein1998}\nocite{Roberts2008}
\nocite{Rubensdotter2003}\nocite{Roberts2011}

\begin{figure}
\ifpreprint
\else
\includegraphics[width=\hsize]{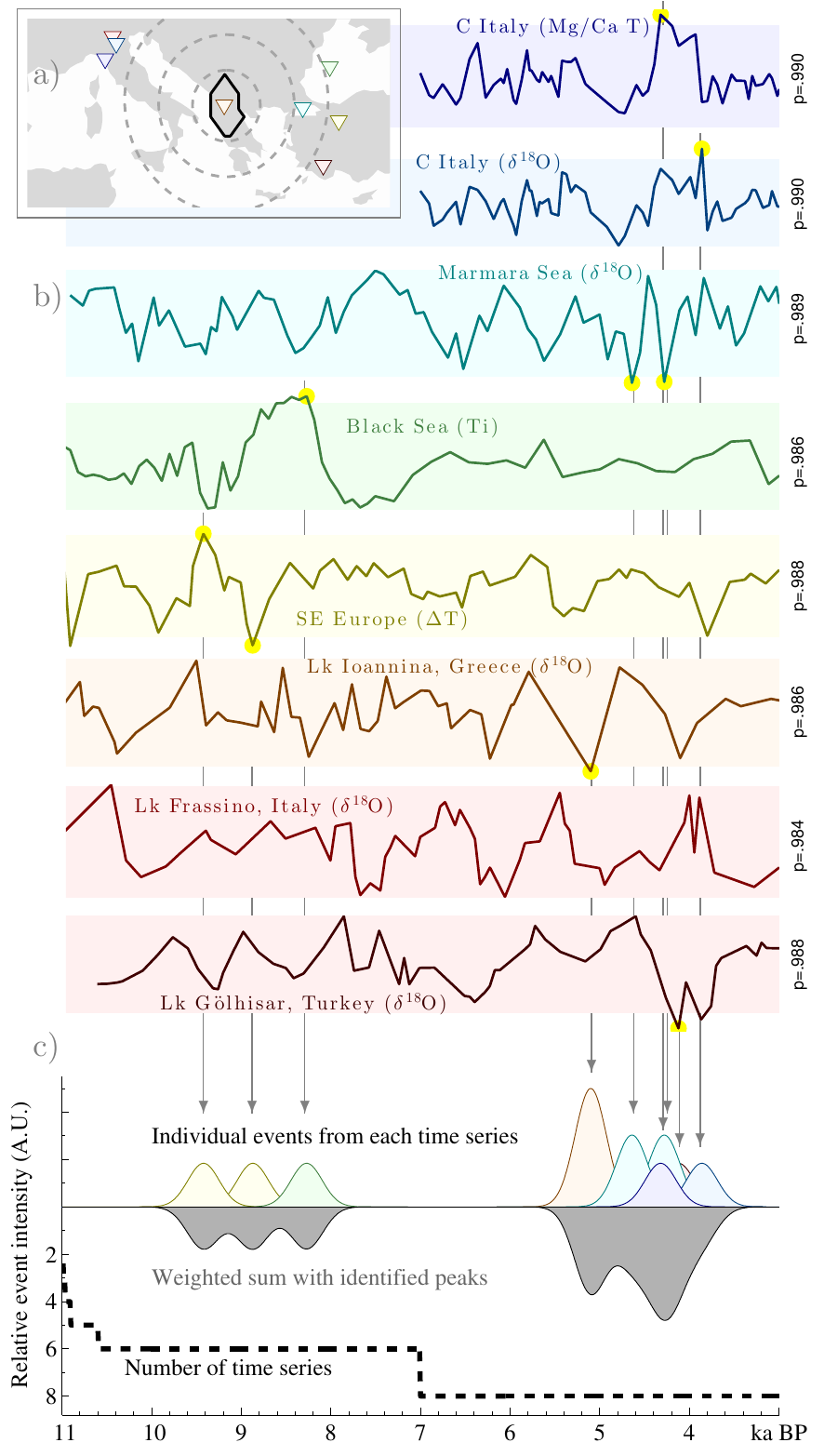}
\fi
\caption{Generation of idealized climate events. 
(a, top) Map with example region and nearby locations (colored triangles) where long-term and high-resolution palaeoclimate information is available;  concentric circles indicate the distance from the simulation region.  (b)~Associated palaeoclimate time series, detrended and normalized; for each time series, the confidence level $p$ is indicated and visualized as the width of the background shading. Climate extremes are identified outside this confidence interval and are highlighted. (c, bottom) Contribution of individual extreme events from palaeoclimate time series (weighted by distance and color-coded as above) to the probability distribution of event occurrence in the simulation region.  The dotted line shows the number of time series contributing to the event generation for this region.}
\label{fig:eventintegration}
\end{figure}

From the global dataset, 26~time series are located in or near our focus area western Eurasia (\reftab{tab:proxies}). For these time series we analyzed the non-cyclic event frequency according to the procedure in \citet{Wirtz2010}: time series were detrended with a moving window of 2000~years and smoothed with a moving window of 50~years, then normalized (\reffig{fig:eventintegration}b).  Events were detected whenever a time series signal exceeded a confidence interval with threshold $p=1-1/n$, where $n$ is the number of data points \citep{Thomson1990}, and where each event is preceded or followed by a sign change in the time series.

For each simulation region, events from spatially overlapping or nearby proxy locations were used to construct an aggregated event time series specific to this region (\reffig{fig:eventintegration}a--c):  (1)~a Gaussian filter with $\tau=175$\,a (corresponding to the dating uncertainty in many records) standard deviation was applied to each event; (2)~the distance of the proxy location to the simulation region was used to assign exponentially decreasing weights to each event; (3)~all event densities were summed and filtered with a running mean with window $\tau$; (4)~all events representing single anomalies above the mean event density were used for further analysis, irrespective of their magnitude. On average, seven events were detected per time series.

From this analysis, we obtain a pseudo-realistic data base of spatially and temporally resolved climate events.  By using this idealized  approach, we show a way to overcome issues raised by, e.g., \citet{Schulting2010} on chronological resolution and spatial representation problems of individual records.  In our study, we consider the impact of the number of extreme events and their spatial patterning rather than single chronologies. This idealized---or potential---climate events database thus allows us to go forward with analyzing the climate-human relationship while the reliability of individual palaeoclimate reconstructions is still being questioned.

\subsection{Global Land Use and technological Evolution Simulator}
The Global Land Use and technological Evolution Simulator (GLUES, \citealt{Wirtz2003gdm,Lemmen2010,Lemmen2011}) 
was developed to study how differences in cultural trajectories at the region scale can be attributed to the specific adaptation of local societies. GLUES mathematically resolves the dynamics of local human populations' density and characteristic sociocultural traits in the context of a changing biogeographical environment.  One of the characteristic traits represents available technologies, a second one the share of farming and herding activities, and the third one the number of established agropastoral economies.  Trait characteristics adapt according to a growth-benefit gradient dynamics \citep[e.g.][for ecological applications]{Smith2011}. Traits are further exchanged between simulation regions by information dispersal or migration. The model is described in more detail in the supporting online material.

Our climate event data base is used to impose sudden reductions ($f$) of the land utility whenever a climate event occurs in a simulation region.  If a local simulated population experiences reduced food yield during a climate event, people attempt to mitigate the crisis by moving to adjacent regions.  In this respect, climate events act as a stimulus to migration in the model.  If outmigration is not possible, population density declines; with severe population decline, also some of the technologies are lost \citep{Lemmen2010rgzm}. 

\begin{figure}
\ifpreprint
\else
\includegraphics[width=\hsize]{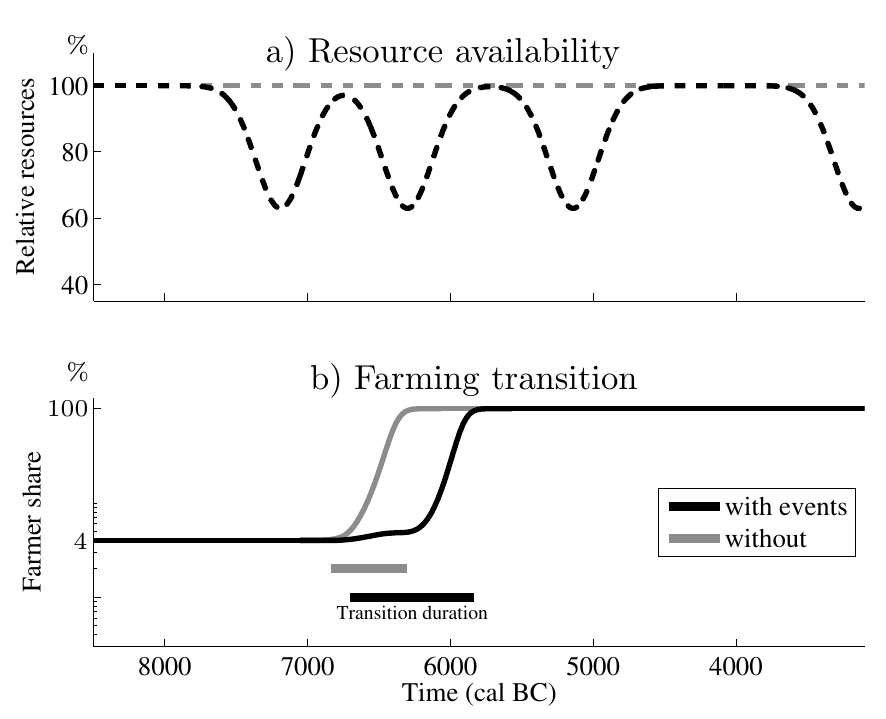}
\fi
\caption{Simulated climate events and evolution of farmer share for an example region. (a, top)  Declining resource availability at times where the palaeoclimate record indicates climate extremes. (b, bottom) Transition from hunting-gathering to agropastoralism in the simulation without (grey) and with (black) climate events.  Both the timing of the transition as well as the duration of the transition (defined as the time needed in the simulation to increase the farmer share from 4.1\% to 95\%) are prolonged by the introduction of climate events in this example region.}
\label{fig:trajectory}
\end{figure}

The simulation is started at 9500\,sim\,BC\footnote{we use the age scale `simulated time BC' (sim\,BC) to distinguish between empirically determined age models (cal\,BC) and the simulation time scale}. All of the 685~biogeographically defined regions are initially set with farming activity at 4\% and established agropastoral communities at 25\%, what represents a low density Mesolithic technology population and a broad spectrum foraging lifestyle with low unintentional farming activity.  We define a local population as Neolithic when the share of agropastoralists is larger than the share of foragers---regardless of its technology, economic diversity, or population density (\reffig{fig:trajectory}).  Experiments are performed with different impact strength ($f$) of extreme events on land utility to assess climate related sensitivities; here we compare the simulation without climate events ($f=0\%$) to simulations where the utility reduction was $f=10\%$ to $100\%$; results are shown for  $f=40\%$, a value that represents a moderate but not excessive impact.

\subsection{Reference data}
%
Our reference is a sub-set of the comprehensive data collection by \citet{Pinhasi2005}, who used site data provided by the United Kingdom Archaeology Data Service, the Central Anatolian Neolithic e-Workshop (CANeW), the radiocarbon CONTEXT database, and the Radiokarbondaten Online (RADON) database.  In their compilation, \citeauthor{Pinhasi2005} included only sites with small dating uncertainty ($\le200$\,a) and report dates based on calibration of original $^{14}$C measurements with CalPal~2004.  The data set contains dates distributed across Western Eurasia and was recommended by its authors for comparison with simulation studies (see \reffig{fig:timingref} for dates and locations).  For numerical comparison with the simulation, we calculate the timing of the Neolithic onset from the radiocarbon date of all sites located within each simulation region.  As there is no standard procedure for calculating such an area-averaged onset \citep{Mueller2000}, we us the average date.

\section{Results}

In the reference GLUES simulation including climate variability, farming originates in the Fertile Crescent and southeast Europe in the 7th millennium sim\,BC and penetrates into Europe in a northwest direction (\reffig{fig:timingref}). By 3500\,sim\,BC all of continental Europe has converted to farming as the predominant subsistence style.  

\subsection{Expansion of agropastoralism}

The initial development progresses slowly and at a low level. It begins during the 67th century~sim\,BC in the Levant and Greece\footnote{We use current geographic names to refer to the simulation regions}, followed by the central Balkan (66th century). From these centers, farming spreads to the western Black Sea coast during the 62nd century, and is present throughout the entire Balkan, Anatolia, and Mesopotamia by the 59th century. In these areas, the northward spread of farming stagnates, while farming activity intensifies up to the 56th century on the Balkan, and while farming expands to Italy and the eastern Black Sea coast. 

From the 55th to the 59th century sim\,BC, a rapid expansion through central Europe---or what is archaeologically seen as the Linearbandkeramik area---is simulated, such that farming is the major subsistence style throughout central and southeastern Europe, including the the northern Black Sea coast.  From the 46th century, farming emerges along the Baltic Sea coast, and has become the dominant subsistence style throughout the North European plains, Poland, and Denmark by the 42nd century. In France and England, first farming is evident from the 39th century. 

Independently, farming also originates in North Africa at the Gibraltar Strait in the 61st century sim\,BC, from where it spreads into Morocco and the Iberian peninsula, penetrating southern Spain and Morocco by the 54th century.  The expansion into northern Iberia stagnates until the 49th century; and slowly connects along the Mediterranean coast to the Fertile Crescent expansion branch by the 46th century.  All of of the Iberian peninsula has converted to farming by the 42nd century.  By the 34th century, all of continental western Eurasia and England rely almost exclusively on farming.  An animation showing the expansion of agriculture into Europe is shown as a supplementary Movie~S1.

While the simulation captures the eastern route of the Neolithic into Europe (via Greece, the Balkan, Hungary, then north- and westward, see \citealt{Rasse2008} for an overview of routes) quite well, it fails to simulate the Mediterranean route (Cyprus, Sicily, French coast, then northward); this was attributed by \citet{Lemmen2011} to the lack of sea transport in the current model.   A western route into Europe is suggested by the model emanating from the Strait of Gibraltar; apart from some zooarchaological evidence \citep{Anderung2005}, this route has not been confirmed by archaeology \citep{Gronenborn2009tcc,Rasse2008}.

\begin{figure*}
\centering
\ifpreprint
\else
\includegraphics[width=.9\hsize]{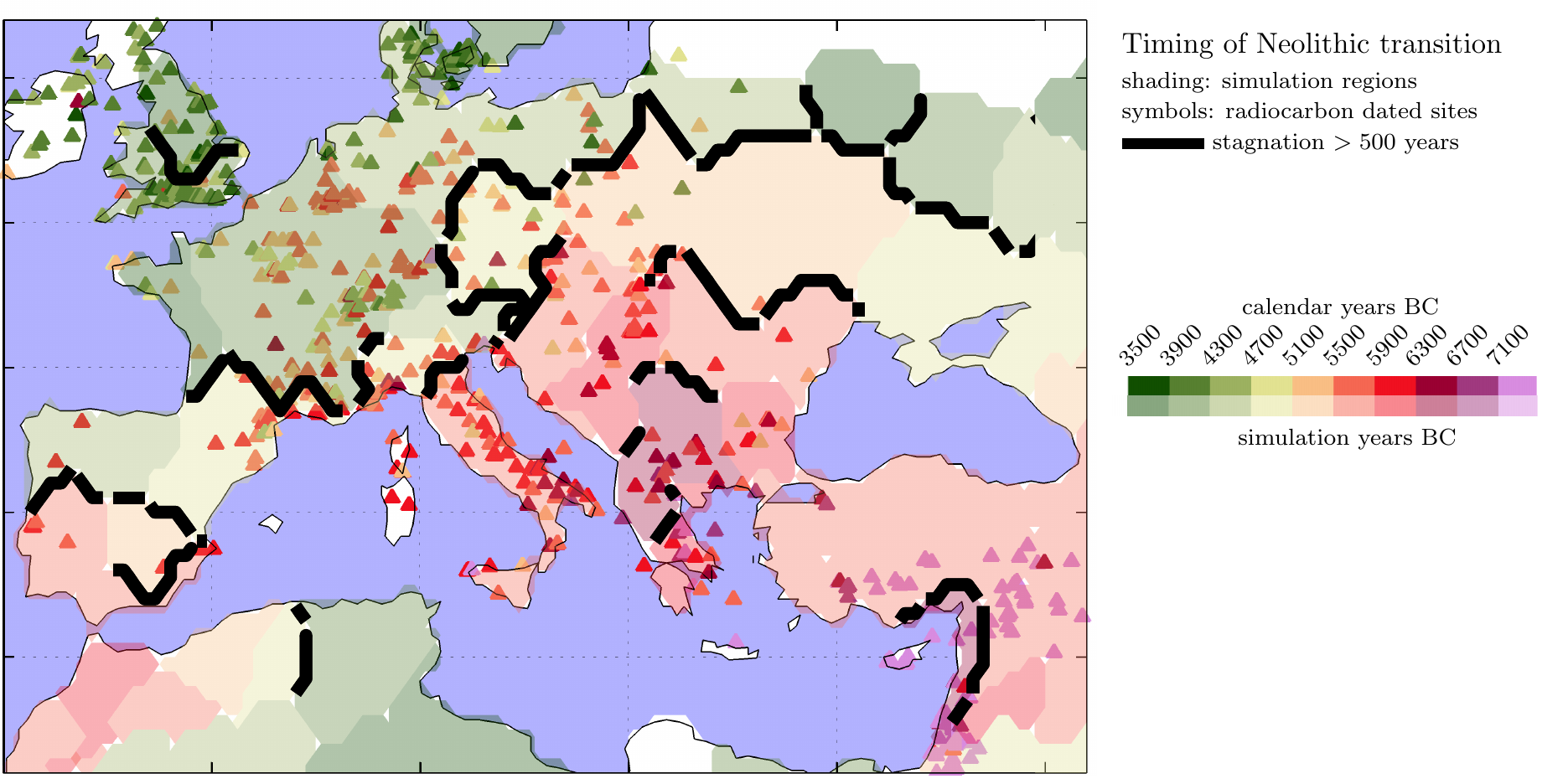}
\fi
\caption{Timing of the transition to agropastoralism in Western Eurasia.  The simulated transition (background pastel shading) is contrasted with the radiocarbon ages of Neolithic sites from  \citet[][solid color triangles]{Pinhasi2005}.  Bold lines indicate regional stagnation periods where the onset lag between neighboring regions is at least 500~years.}
\label{fig:timingref}
\end{figure*}
%
The regional timing of agropastoralism is contrasted with the median radiocarbon dates of Neolithic sites compiled by \citet{Pinhasi2005} (\reffig{fig:timingref}). From this synoptic, time-integrated perspective, the simulated centers of agropastoralism in the Fertile Crescent, in northern Greece and at the Strait of Gibraltar are evident, as well as the southeast to northwest temporal gradient of the Neolithic transition.  The general pattern of the simulated transition resembles the pattern that can be seen from the radiocarbon dates; locally, many dates deviate from the large-scale simulation or disagree with each other within a simulation region.  

\begin{figure}
\ifpreprint
\else
\includegraphics[width=1\hsize]{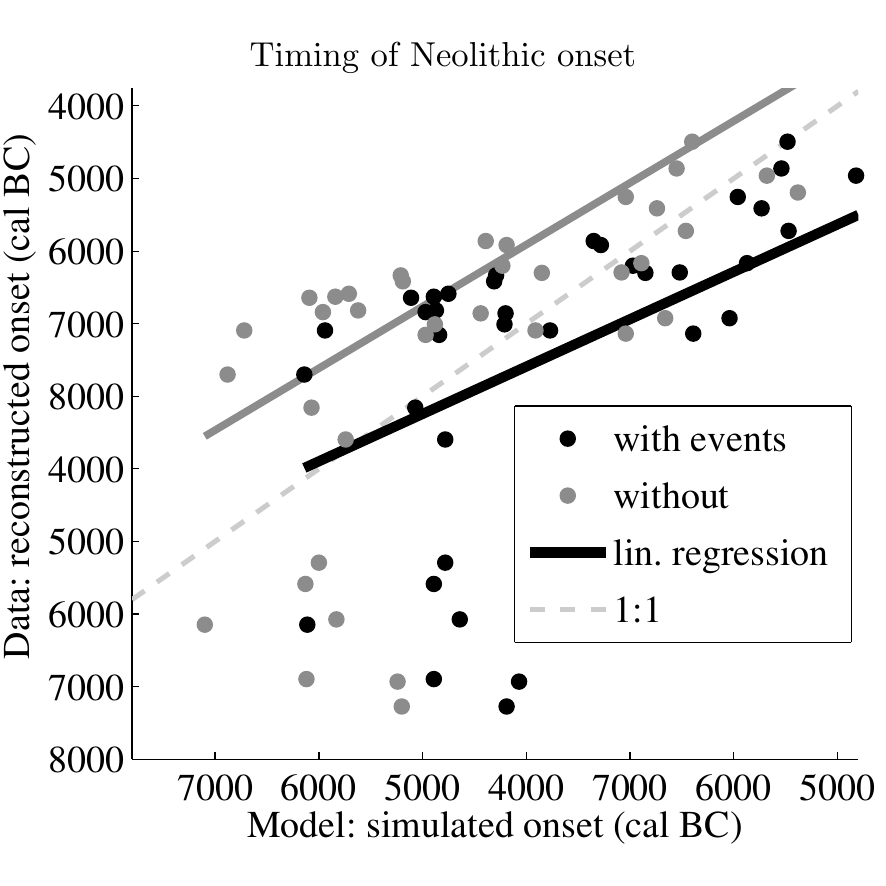}
\fi
\caption{Reconstructed timing of agropastoralism versus simulated onset using a constant or a fluctuating climate forcing
(grey and black circles, respectively). The lower left cluster of outliers represents Fertile Crescent and Anatolian founder regions.}
\label{fig:transition}
\end{figure}

Overall, the model skill visualized in \reffig{fig:transition} indicates that recorded variability both between and within European regions can be sufficiently well reproduced by the combination of migration and endogenous dynamics as formulated in GLUES.  There is a significant correlation between the hindcasted onsets and the reconstructed onset in both the simulation with and without climate events ($r^{2}=.37, r^{2}=.43, n=39, p>.99$, respectively). 
The simulation with climate events hindcasts the onset for most regions 162~years later than the average reconstructed onset; the mean model bias is
$\approx 400\pm1000$~years; the difference to the simulation without climate events (median onset 282~years earlier than the reconstruction) is statistically not significant.

On average, climate events delay the onset by $461\pm179$~years.  Small delays (minimum 65~years) occur in regions where climate events are temporally separated from the transition, while long delays (maximum 1150~years) occur in regions where several climate events occur shortly before and in the initial phase of the transition.  Climate events during the transition tend to prolong the duration of the transition by 50--100~years but this trend is not statistically significant. 





\section{Discussion}
\subsection{Stagnation lines in model and archaeological evidence}
%
The Global Land Use and technological Evolution Simulator is able to hindcast a realistic spatiotemporal pattern of the introduction of farming and herding into Europe between 7000 and 3500\,sim\,BC. Simulated transitions towards agropastoralism compare well to a large dataset of radiocarbon dated Neolithic sites. In the simulation, as well as in the data, agropastoralism did not expand uniformly, but rather in periods of rapid spread interrupted by periods of spatial stagnation---but local intensification.  This rapid Neolithization, for example, applies to the expansion from Greece to the central Balkan in the 67th century sim\,BC, which is followed by a 400\,a period of relative stagnation. A very similar pattern is hindcasted for the Linearbandkeramik-like Neolithization in the 55th and 54th century cal\,BC, and for the relative stagnation before the onset of a Funnelbeaker-like cultural complex further north, in the 49th to 46th century. Other stagnation lines identified in the model occur around the Levant and along the Pyrenees, and divide the southern from the northern Iberian peninsula.

These stagnation periods were archaeologically recognized in \citeauthor{Guilaine2003}'s~(2003) `hypoth\`ese arythmique'.  Based on this hypothesis,  \citet{Rasse2008} identified stagnation lines by analyzing isolines in the European Neolithic radiocarbon record: he also found long duration stagnation lines running through Iberia and the Pyrenees, and one crossing Anatolia; shorter stagnation lines were identified in Greece, one bisecting the Balkan, another running along the Carpathian mountains, and one along the Alps.  Extending the earlier work, \citet{Schier2009} emphasized a major stagnation line along the northern edge of the European loess belt, separating the Linearbandkeramik and Funnelbeaker cultures between 5100 and 4400~cal\,BC. 

Dispersal models of the European Neolithic create spatial structure in the simulated Neolithization pattern from geographic and topographic constraints  \citep[e.g.][]{Davison2006,Ackland2007,Galeta2011}. We add to this infrequent climate excursions emerging as sudden decreases of the natural productivity and diffusion rates of traits and people.  This diffusion not only depends on the topography and geography, but again on (simulated) people and their technology. Compared to those dispersal studies, our model draws a more heterogeneous picture of the regional transitions to agriculture; it is the only model which is capable of representing stagnations in the expansion.
We are thus one step further in establishing a simulation of the deterministic pattern of the Neolithic in Europe.  Where this deterministic pattern fails to reproduce archaeological information, we can in the future more precisely identify times and regions where human agency  played a dominant role.  

Before the discrepancies can be exploited scientifically, several caveats about the simulation have to be considered: (1)~the model deviation is larger than the uncertainty associated with radiocarbon dates for most sites.  (2)~each simulation region covers a large area which cannot be fully represented by individual sites, such that the scale difference introduces an additional comparison uncertainty; (3)~there is a  a spatial bias in the data: the number of sites per simulation region varies, and there is good coverage along the transect from the Levant to northwestern Europe, but few or no information on eastern Europe, on the Iberian peninsula, and in North Africa.   

\subsection{Role of climate events}
Reconstructed climate events episodically depressed population density and to a lesser extent technological stages. The impact of events on the timing of the Neolithic simulation is largest whenever they occurred before or during the first stage (between 10--50\% agropastoralists) of a local transition:  In 29 regions (out of the 53 simulation regions in Europe), a climate event occurred during the transition. On average, the transition was delayed by 50--100~years (w.r.t.\  a simulation without events).  In three cases, a regionally emerging agropastoral life style reverted back to hunting and gathering during a climate event. 

Not everywhere, however, where a climate event coincides with the transition, there is a  delay; and vice versa, there are delays in regions which did not experience climate events during the transition to agriculture, such as in most parts of western central Europe. All individual regions are closely connected by neighbor and remote trade relations: impacts, such as those of climate events, may have had far-reaching consequences beyond the locally affected region \citep{Weninger2005}.

As evident from comparing this study with the one by \citet{Lemmen2011}, climate fluctuations hindered the development in Greece and the Balkan 
 (which is now more realistic), while the temporal gaps between the central Balkan, the Linearbandkeramik  culture, and the Funnelbeaker regions become more pronounced, which is also more consistent with archaeological evidence \citep{Pinhasi2005,Barker2006,Schier2009}.  Compared to the field evidence, the simulation shown here is too late in reproducing the transition to farming in France.  Though migration is not enhanced by climatic triggers in continental Europe, it does increase so for Great Britain, where the transition occurs as early as France thanks to increased immigration. 

The differences we obtain between simulated timing and the radiocarbon age of related sites  (Figures~\ref{fig:timingref}) indicate a partial improvement over the simulation without climate events shown by \citet{Lemmen2011}:  the overall bias of the simulation increased slightly (the event forcing led to an overall delay of the onset), but the variability in the data could be redrawn more precisely. Despite the non-significant regressions between reconstructed and hindcasted onsets and durations of the Neolithic, the scatter alignment is close to the 1:1~line---apart from a bias in the onset hindcasts for the Levante regions (\reffig{fig:transition}); neither regionally clumped chronologies nor model assumptions can be expected to be quantitatively precise. 

Overall,  uncertainties in the simulated timing, the radiocarbon dates, and in the regional up-scaling  are yet so large, that the differences between the simulation with and without climate events are not statistically significant for Europe.  Yet, the reconstruction of climate events improved the hindcasted Neolithic transition, both in timing and regional transition durations. On the one hand, this points to an effective albeit limited sensitivity of societal dynamics to environmental shifts. On the other hand, this result supports the notion of a large resilience of cultures, just like \citet{Grosman2003} found for the Late Natufians. Agency, the capacity of cultural complexes to adapt and preserve their life-styles still appears as the dominant model for understanding past human ecodynamics.  

This short paper is only a start in the direction of tackling the climate-culture link on large scales, and it gives a perspective of what could be achieved by bringing palaeoclimate data, cultural modeling, and chronologies together.  The idealization of climate events we present here needs to be discussed and evaluated further, likewise the (few) assumptions made by the model. In this attempt, the increasing availability of local chronologies will stimulate regional-scale simulation studies for Europe and other continents.

\section{Conclusion}

We presented a spatially explicit mathematical model of the
Neolithization of western Eurasia from 7000\,BC to 3500\,BC.  Our model
incorporates endogenous sociotechnological dynamics, as 
represented by the adaptation of characteristic population traits and
their interaction with demographics and changing environments. 
The simulation improved in hindcasting the Neolithic expansion after integrating quasi-realistic climate events, especially with regard to stagnation phases visible also in
field data. This outcome helps to confine the relevance of climate
variability in influencing past human ecodynamics and suggests a
dominance of endogenous over exogenous control factors.

\section*{Acknowledgments}

We acknowledge the financial support for C.L.\  by the German National Science Foundation (DFG priority project Interdynamik~1266), and for both authors by the PACES programme of the Helmholtz society (HGF).  We thank the editor and anonymous reviewer for thoughtful comments on an earlier version of this manuscript. This work would not be possible without the many data contributors who made their data publicly available or provided it to us for analysis, thank you. 

\section*{Supporting material}
In the supporting online material, a movie of the expansion of agriculture (Movie~S1), the full table of globally distributed palaeoclimate time series (Table~S1), and a more detailed description of GLUES are available. The simulated data have been permanently archived on and are freely accessible from PANGAEA (Data Publisher for Earth \& Environmental Science) as a netCDF dataset with reference \nobreak{doi:10.1594/PANGAEA.779660}.  GLUES is free and open source software and can be downloaded from http://glues.sourceforge.net.

\bibliographystyle{elsart-harv}
\bibliography{mendeley-manual}
\end{document}